\documentclass[pre,twocolumn,groupedaddress,showpacs]{revtex4}
\usepackage{graphicx,epsfig,amsmath}

\begin{document}

\title{Structural transformations of \\even-numbered \textit{n}-alkanes confined in mesopores}

\author{P. Huber}
\email{p.huber@physik.uni-saarland.de}
\author{V.P. Soprunyuk}
\author{K. Knorr}
\affiliation{Technische Physik, Universit\"at des Saarlandes,
D-66041 Saarbr\"ucken, Germany\\}

\date{\today}

\begin{abstract}
The n-alkanes C$_{12}$H$_{26}$, C$_{14}$H$_{30}$, and
C$_{16}$H$_{34}$ have been imbibed and solidified in mesoporous
Vycor glass with a mean pore diameter of 10nm. The samples have
been investigated by x-ray diffractometry and calorimetric
measurements. The structures and phase sequences have been
determined. Apart from a reduction and the hysteresis of the
melting/freezing transition, pore confined C12 reproduces the
liquid-triclinic phase sequence of the bulk material, but for C16
an orthorhombic rotator mesophase appears that in the bulk state
is absent for C16 but well known from odd numbered alkanes of
similar length. In pore confined C14 this phase shows up on
cooling but not on heating.
\end{abstract}


\maketitle


\section{Introduction}
\begin{figure}[!]
\epsfig{file=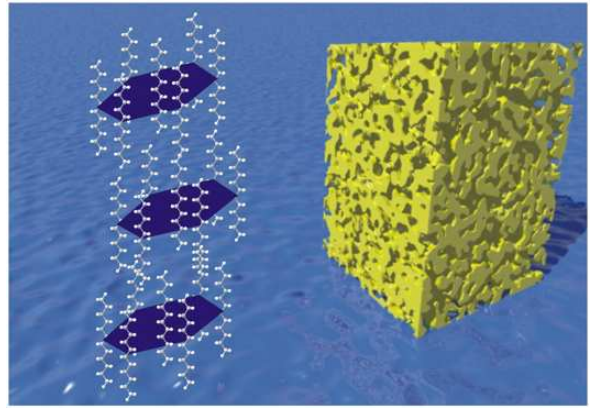, angle=0,
width=0.9\columnwidth}
 \caption{\label{fig1}(Color online)
Schematic view of an n-alkane crystal (left) and a raytracing
illustration of mesoporous Vycor (right).}
\end{figure}

The n-alkanes (C$_{\rm n}$H$_{\rm 2n+2}$, abbreviated Cn) of
intermediate length form lamellar crystals. In the fully ordered
state at low temperatures the odd numbered alkanes exist in the
orthorhombic phase with the molecules perpendicular on the layers
- see Fig. \ref{fig1}, whereas the even numbered alkanes show
phases of lower symmetry, triclinic or monoclinic, with the
molecules tilted away from the layer normal \cite{Dirand2002}.
Close to melting mesophases appear which still have the
translational symmetry of crystals but in which the rotational
degrees of freedom of the molecules about their long axis are
partially or completely disordered \cite{Dirand2002, Sirota
R-phases}, the most prominent being the rotator phase $R_{I}$
(Fmmm, orthorhombic). According to Sirota and Herhold the odd-even
effect in the crystal structures and phase transition temperatures
is closely related to the stability (for n-even$\geq$22 and
n-odd), metastability (n=20,22) or transient appearance
(even-n$\leq$18) of the rotator phase \cite{SirotaHerold}. In
cooling runs a monolayer of the rotator phase already forms a few
K above bulk freezing (an effect known as "surface freezing") at
the liquid-vapor interface\cite{SurfaceFreezing} which then serves
as nucleus for bulk crystallization. The life time of the
transient R phase has been reported to be several seconds for C16
and a few minutes for C18 before the material transforms into the
stable triclinic modification \cite{SirotaHerold,Japsen}.
Analogous observations have been made for emulsified C16 droplets
with a diameter of 33$\mu m$ \cite{nAlkanEmulsionen}. In smaller
droplets (with diameters down to 125nm) there is evidence for a
stable R mesophase for C18 but not for C16. Leaving the
short-lived transient state aside, the melts of C18 and of the
shorter even numbered alkanes directly freeze into the fully
ordered triclinic solid \cite{Dirand2002}. C20 is special in the
sense that it is the shortest even-n alkane that shows a rotator
phase, but only on cooling whereas on heating it melts directly
out of the triclinic phase \cite{Sirota R-phases}.

We have recently started an investigation of alkanes embedded in
mesoporous glasses with pore diameters of 10nm \cite{Huber2004}.
Pore confined C19 shows the phase sequence liquid $L$ - rotator
phase $R_I$ - "crystalline" low-T phase $C$ known from the bulk
system, except for a minority $R_{II}$ state (rhombohedral) in
coexistence with the liquid right at the freezing/melting
temperature and a downward shift of the transition temperatures
$T_{L-R_I}$ and $T_{R_I-C}$. Furthermore both transitions show
hysteresis with respect to heating and cooling. Similar effects on
the transition temperatures are known for many other pore
fillings\cite{Christenson, AlbaSim}, including small molecules
such as He \cite{HeSchmelzpunktred}, Ar \cite{ArSchmelzpunktred},
CO, N$_{\rm 2}$ \cite{CON2Schmelzpunktred}, H$_{\rm 2}$O
\cite{H20Schmelzpunkt}, suggesting that pore confinement favors
less ordered phases with respect to more ordered phases. In the
present paper we examine the behavior of the even numbered
n-alkanes C12, C14, and C16.

\section{Experimental}
Vycor glass (code 7930, Corning Glass Works) with a porosity of
30$\%$ is imbibed with the alkane melts. The structure of the
mesoporous host can be described as a network of 3D randomly
oriented, connected pores with relatively uniform diameter $d \sim
$ 10nm \cite{Levitz1991}. In Fig. \ref{fig1} a raytracing
illustration of the matrix with a somewhat exaggerated tortuosity
of the pores in comparison to real Vycor is depicted. The samples
are mounted in a closed cell that is attached to the cold plate of
a closed cycle refrigerator. They are investigated by means of
standard x-ray powder diffractometry employing coupled $2
\Theta-\Theta$ scans with the Cu K$\alpha$ x-ray beam reflected
from the face of a Vycor tablet. Powder patterns have been taken
as function of temperature $T$ both on cooling and heating, with
$T$-steps down to 1K at phase changes. Recording a powder pattern
($5deg < 2 \theta < 40deg$) took several hours, the waiting time
for $T$-equilibration when changing from one $T$ to the next was
$1/2$h. Thus the experiment only gives information on stable or at
least long-lived structural states. The x-ray diffraction patterns
will be presented as plots of the scattered intensity versus the
modulus of the scattering angle 2$\Theta $ (top axis) and the
modulus of the scattering vector $q$, $q = 4\pi $/$\lambda $
sin($\Theta )$ (bottom axis), where $\lambda $ corresponds to the
wavelength of the x-rays, $\lambda = 1.542\AA$. For C14 and C16
complementary DSC-scans have been taken with a heating and cooling
rate of 0.5K/min.

\section{Results}
\begin{figure}[!]
\epsfig{file=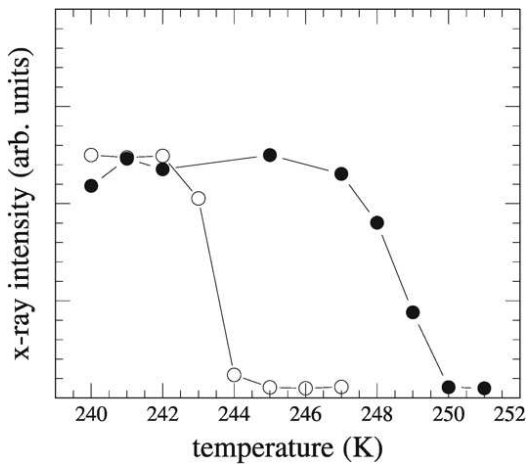, angle=0, width=0.8\columnwidth}
\caption{\label{fig2} The temperature dependence of the integrated
intensity of a group of Bragg reflections of C12 in Vycor, both on
cooling ($\circ$) and heating ($\bullet$).}
\end{figure}

\begin{figure}[!]
\epsfig{file=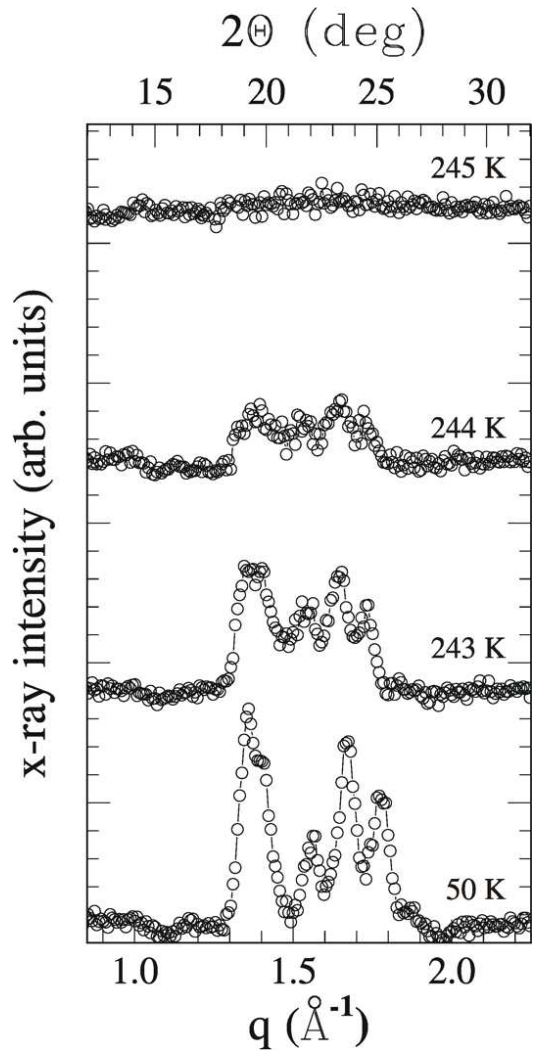, angle=0, width=0.8\columnwidth}
\caption{\label{fig3} Diffraction patterns on C12 in Vycor at
selected temperatures.}
\end{figure}

\begin{figure}[ht]
\epsfig{file=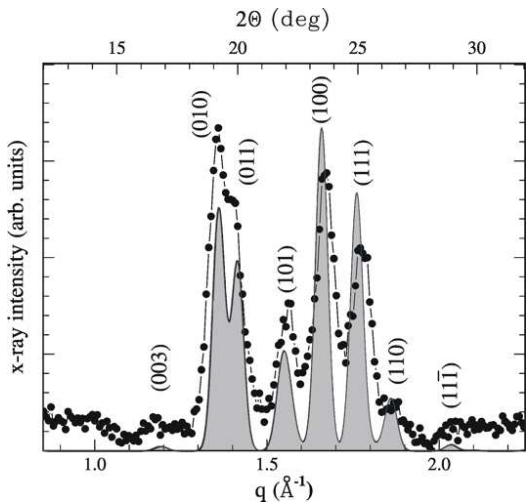, angle=0, width=0.8\columnwidth}
\caption{\label{fig4} The diffraction pattern of C12 in Vycor at
50K in comparison with a powder pattern of bulk C12 (triclinic)
calculated from the structural data of ref. \cite{Nyburg}.}
\end{figure}

Pore confined C12 freezes at $T_{\rm f}$=243K and melts at $T_{\rm
m}$=249K as can be seen from Fig. \ref{fig2} which shows the
$T$-dependence of the integrated intensity of a group of Bragg
peaks both for cooling and heating. The melting temperature of the
bulk system is 263.6K \cite{Dirand2002}. The powder pattern of the
solid regime is shown in Figs. \ref{fig3} and \ref{fig4}. There
are no changes of this pattern with $T$ that suggest a solid-solid
transition. The comparison to the pattern of the bulk triclinic
solid which has been calculated from the structural data of ref.
\cite{Nyburg} indicates that the structure of pore confined solid
C12 is identical to that of the bulk counterpart, apart from a
line broadening due to the finite size of the pore confined
nanocrystallites. Thus the phase sequence liquid-triclinic of the
bulk system is not changed. The coherence length, as extracted
from the width of the Bragg peaks while properly taking into
account the instrumental resolution, was determined to
12($\pm1)$nm.

\begin{figure}[ht]
\epsfig{file=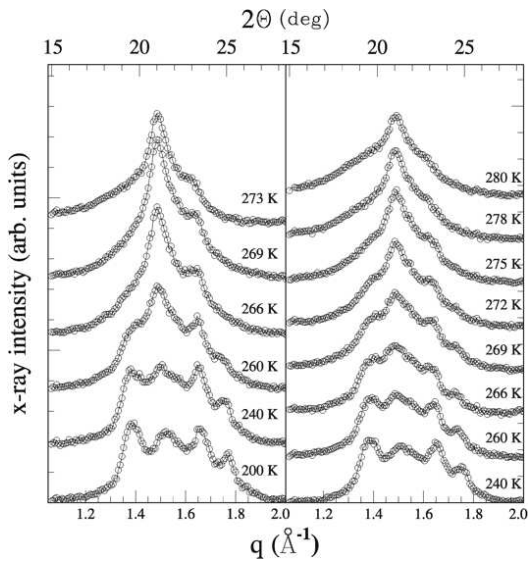, angle=0, width=0.8\columnwidth}
\caption{\label{fig5} Diffraction patterns of C16 in Vycor at
selected temperatures, both on cooling (left) and heating
(right).}
\end{figure}

Fig. \ref{fig5} shows diffraction data on pore confined C16.
Solidification (at 273K) resp. melting (at about 281K) is apparent
from the pertinent changes of the diffraction pattern. In the
liquid state the pattern is dominated by the broad first maximum
of the structure factor whereas in the solid state Bragg peaks
show up. The bulk melting temperature is 291.2K. The pattern of
the solid state starts out as a two-peak-pattern, with the second
peak having the form of a shoulder sitting on the high-q wing of
the first peak, and changes to a four-peak pattern at lower $T$.
The crossover temperature is about 260K on cooling and 270K on
heating.

The two components of the two-peak-pattern are centered at 1.49
and 1.63{\AA}$^{ - 1}$. This is what is expected for $q$-values of
the principal in-plane reflections (110) and (200) of the
orthorhombic rotator phase $R_{I }$\cite{Sirota R-phases}. At
lower $T$ the pattern approaches but does not really reach the
pattern of the triclinic state. Three of four stronger diffraction
peaks are consistent with the reflections (010)/(011), (100), and
(111) of the triclinic structure, the peak centered at
1.49{\AA}$^{ - 1}$ is not. This peak is still at the position of
the strong (110) reflection of the $R_{I}$ phase and is also
somewhat broader, due to an admixture of the weaker triclinic
(101) reflection. Obviously pore confined C16 settles at low $T$
in a state of triclinic-orthorhombic coexistence. One expects of
course that the $R_{I}$ minority component eventually transforms
at low $T$ into the ``crystalline'' herringbone phase, but the
intensities of the extra reflections of this phase are too weak to
be detected.

Thus the phase sequence is liquid-$R_{I}$-$R_{I}$/triclinic
coexistence. The stable intermediate $R_{I}$ phase is an extra
feature of the pore confined system. The sequence as such is
reversible with respect to cooling and heating but the transition
temperatures show thermal hysteresis.

\begin{figure}[htbp]
\epsfig{file=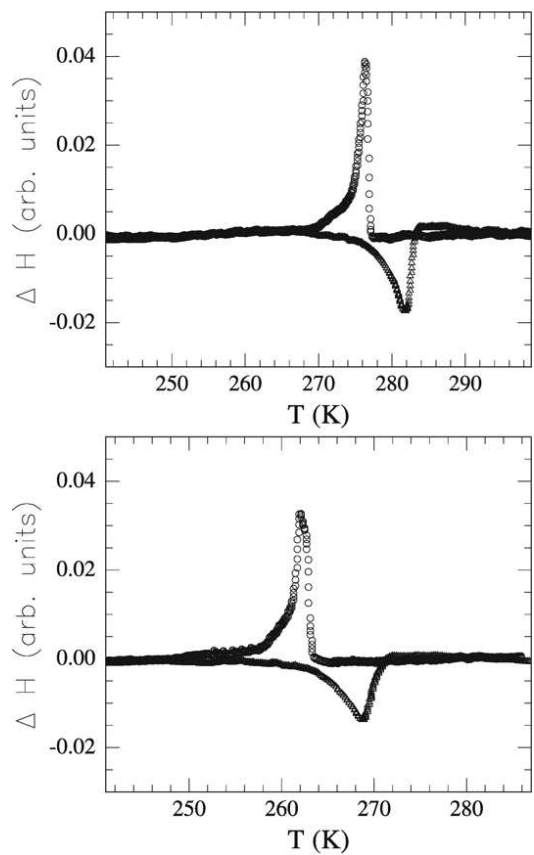, angle=0, width=0.8\columnwidth}
\caption{\label{fig6} Differential scanning calorimetry data on
C16 (upper panel) and C14 (lower panel), both on cooling ($\circ$)
and heating ($\triangle$).}
\end{figure}

The DSC data (Fig. \ref{fig6}) confirm the values of $T_{\rm f}$
and $T_{\rm m}$ of the diffraction experiment. On the other hand
there is no evidence in this data for a partial $R_{I}$-triclinic
transformation. This may be due to the fact that the latent heat
of this transformation is distributed over a broad $T$-interval,
which is in fact suggested by the diffraction data.

\begin{figure}[!]
\epsfig{file=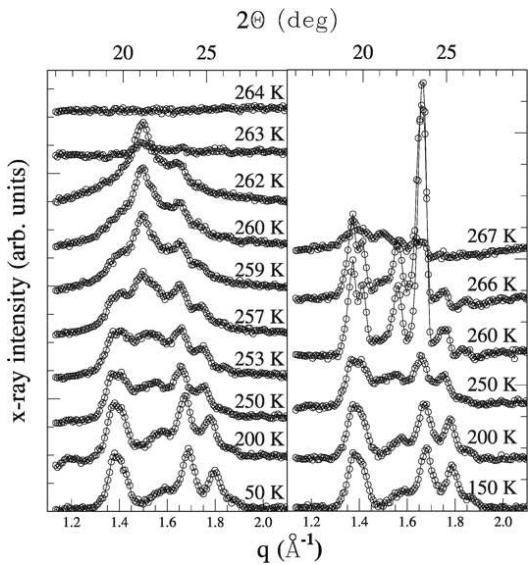, angle=0, width=0.8\columnwidth}
\caption{\label{fig7} Diffraction patterns of C14 in Vycor at
selected temperatures, both on cooling (left) and heating
(right).}
\end{figure}

The diffraction results (Fig. \ref{fig7}) on pore confined C14 are
in a sense intermediate to those on pore confined C16 and C12. On
cooling the two-peak profile of the rotator phase $R_{I}$ appears
at T$_{\rm f}$=263K (the bulk melting temperature is 279K), below
259K the systems shows $R_{\rm I}$-triclinic coexistence that
eventually below about 250K purifies into the triclinic single
phase state. On heating the $R_{\rm I}$ phase is suppressed and
the pore filling melts directly from the triclinic state at 267K.
Nevertheless the behavior on heating is peculiar. At the end of
the cooling run the triclinic reflections are relatively broad,
the (010)/(011) doublet is for instance not resolved. This
suggests that the triclinic crystallites are small in size and/or
heavily strained, perhaps due to the presence of $R_{\rm I}$
residues. During the heating cycle the peaks sharpen and stay so
up to the melting point at 269K. The crystallites obviously grow
in size and built-in strains relax. This ripening occurs at about
260K which is roughly the temperature of the $R_{\rm I}$-triclinic
phase transformation of the cooling cycle, that is in a $T$-range
of appreciable thermal agitation. Thus the phase sequence is
liquid-$R_{\rm I}$-triclinic on cooling and triclinic-liquid on
heating, and one has to distinguish between the quenched and the
annealed version of the triclinic state. With respect to the
appearance of the $R_{\rm I}$ phase, the sequence is irreversible.
When the heating cycle is stopped prior to melting and the sample
is cooled down again, the reflections remain sharp.

As for C16, the diffraction and the DSC data agree on the values
of $T_{\rm f}$ and $T_{\rm m}$ (Fig. \ref{fig6}). The long wing of
the freezing anomaly of the DSC experiment that extends down to
about 250K may be related to the gradual $R_{I}$-triclinic
transformation.

In the $R_{I}$ phase of pore confined C14 and C16 only the two
strongest in-plane reflections could be detected. The (00l)
layering reflections are absent, in agreement with the situation
for the pore confined odd-numbered alkane C19. This means that the
lamellar arrangement is suppressed or at least heavily perturbed.
Note that already mean square displacements of the molecules in
the direction normal to the lamellae of 1 or 2{\AA} wash out the
modulation of the electron density in this direction. In the
triclinic state with tilted molecules, however, the mere existence
of the mixed reflection (111) proves that the lamellar arrangement
is still intact.

\section{Discussion}

In the bulk state of even numbered alkanes, stable rotator phases
no longer exist for $n<20$. In the pore this stability limit is
shifted to $n<14$. In fact pore confined C14 shows exactly the
phase sequence of bulk C20, with the rotator phase existing on
cooling but being inaccessible on heating. The question of course
arises whether the $R_{I}$ mesophase and the pore confined state
in general is stable, metastable or long-lived instable. From an
experimental point of view the diffraction patterns do not change
with time over isothermal waiting periods of several days.

\begin{figure}[!]
\epsfig{file=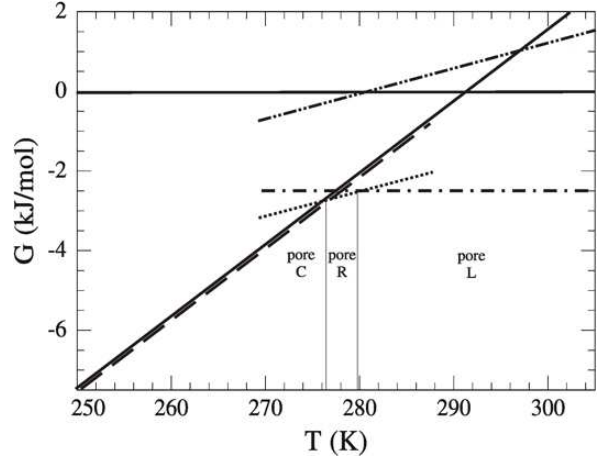, angle=0, width=0.9\columnwidth}
\caption{\label{fig8} Schematic diagram of the Gibbs energy G of
C16 as function of temperature, both for the bulk and the pore
confined state, relative to the bulk liquid ($G_{L}^{bulk}$=0).
The bulk $C$-phase is shown by a solid line, the hypothetical bulk
R phase as dash-dot-dot line, the pore liquid as dash-dotted line,
the pore $C$ phase as dashed line, and the pore $R$ phase as
dotted line.}
\end{figure}

In order to explore the question of stability we refer to the
phase diagram of Fig. \ref{fig8} that show the Gibbs energy
$G=H-TS$ of the liquid $L$, the rotator $R$ and the triclinic
crystalline phase $C$ of C16 as function of $T$, both for the bulk
and the pore confined state. We assume that the enthalpies $H$ and
the entropies of the three phases do not vary with $T$ and the
entropy of a given phase is the same in the bulk and the pore
confined state. $G_{\rm L}^{\rm bulk}$ serves as zero reference.
$G_{\rm C}^{\rm bulk }$ is known from experimental data on the
melting temperature and the heat of fusion of C16. $G_{\rm R}^{\rm
bulk }$ is constructed from an extrapolation of the transition
temperatures and latent heats of other alkanes (even-n, n$ \ge
$20, and odd-n) that do show a $R_{I}$ rotator phase
\cite{Dirand2002}. The Gibbs energies of the pore confined phases
are displaced with respect to their bulk counterparts, $G_{\rm
A}^{\rm pore}$=G$_{A}^{\rm bulk}-\Delta _{A}$, $A=C,R,L$. For the
liquid state, $\Delta_{L}$ is positive, the pore liquid is stable
with respect to the bulk liquid outside the pores as can be seen
from the fact that a drop of liquid is sucked into the pores, due
to the attractive interaction between the molecules and the pore
walls. The pore solid also benefits from this interaction but it
has to pay an extra price in form of strains, defects, grain
boundaries that are required to match the solid to the pore
geometry. If we assume that C12 stays in the pores upon
solidification into the $C$ phase for thermodynamic reasons and
not due to kinetic barriers for the extrusion of the solid,
$\Delta_{C}$ should be lower than $\Delta_{L}$, but still
positive. $\Delta_{L}$ is unknown for the alkanes of the present
study, but a rough idea can be obtained from C8 where the vapor
pressure at the melting point is still large enough to allow us a
measurements of an adsorption isotherm which gives direct
information on $\Delta _{L}$, $\Delta _{L} \approx $2500Jmol$^{ -
1}$. Using this value, $\Delta _{C}$ can then be estimated to be
of the order of 100Jmol$^{ - 1}$ from the shift of the $C-L$
melting transition of C16 upon pore confinement, $T_{C -
L}^{pore}$-T$_{C - L}^{\rm bulk }$= ($\Delta
_{C}-\Delta_{L})$/S$_{C - L}$, S$_{C - L}$ is the entropy change
at the $C-L$ transition. This is further evidence that the pore
confined $C$ solid is stable, but only marginally so.

Whether the $R$ solid is stable in the pores depends on the value
of $\Delta _{R}$. It is reasonable to assume that $\Delta _{R}$ is
intermediate to $\Delta _{C}$ and $\Delta _{L}$. It turns out that
$\Delta _{R}$ has to be practically equal to $\Delta _{L}$ in
order to arrive at a stable $R$ phase in the pores, but its
$T$-range of existence is very small. See the phase diagram of
Fig. \ref{fig8} that is based on the choice $\Delta _{C}$=100,
$\Delta _{R}=\Delta _{L}$=2500 in units of Jmol$^{ - 1}$.
Analogous considerations of C14 and C12 show that the pore
confined $R$ phase of these alkanes cannot be stabilized for any
$\Delta _{R} \le \Delta _{L}$. In case $\Delta _{R}$ and $\Delta
_{L}$ are equal there should be no shift of the $R-L$ melting
temperature by pore confinement and in fact the example of C19
which melts from the $R$ phase shows that this is almost so
\cite{Huber2004}. (In bulk C16 the $R-L$ transition is not
accessible because of the appearance of the $C$ phase, see Fig.
\ref{fig8}). The equality of $\Delta _{R}$ and $\Delta _{L}$
suggests furthermore that the $R$ solid can be easily matched to
the pore geometry, very much like the liquid, with little extra
energy cost, due to the high level of intrinsic disorder. Indeed
rotator phases are occasionally called ``plastic'' since they can
be extruded by pressures much smaller than required for completely
ordered phases \cite{Michils1948}.

The discussion in terms of $G,T$ phase diagrams gives hints as to
the stability of the phases involved but fail to reproduce the
$T$-width of the $R$-phase of C16 let alone the appearance of the
R phase of C14 on cooling and the $R-C$ coexistence of C16 at low
$T$. We think that structural gradients across the pores have to
be considered. The molecule-substrate potential decays with the
distance from the pore walls, hence the state of the pore filling
next to the pore walls is different from the state in the pore
center, including the possibility of a radial arrangement of
coexisting phases. For the solidification of Ar in porous glass we
could explain most of the pertinent experimental observations in
terms of a simple thermodynamic model that is based on the idea
that the solidification takes place in the pore center, but that
there is a shell of liquid between the solid core and the pore
walls, the thickness of which decreasing slowly with decreasing
$T$ \cite{Wallacher2001}. Such a model could also explain the
$R-C$ coexistence in pore confined C16, with the $C$ phase in the
pore centre surrounded by a matching layer of the $R$ phase. The
fact that the diffraction pattern of C16 does not change anymore
below about 260K simply means that this state is frozen-in below
this temperature, that the thermal energy is no longer sufficient
to drive the phase transformation.

As to the appearance of the $R$ phase of C14 on cooling we cannot
present a convincing argument. One could think of undercooling in
the sense that the $C$ phase cannot nucleate in the pore center
because of a lack of nucleation sites or refer alternatively to
the transient $R$ phase of ref. \cite{SirotaHerold} as vehicle for
the further growth of a metastable $R$ phase rather than for the
growth of the stable $C$ phase.

In more general terms one can argue that the tortuous character of
the pore network of Vycor and rough pore walls act as sources of
random strain fields that stabilize the disordered $R$ phase and
thereby lower the temperature of the $R-C$ transition.
It is conspicuous that the rotator phase that appears in the pore
confined state is the $R_{I}$ phase, the prototypic rotator phase
of the odd numbered bulk alkanes, and not one of the tilted
rotator phases of the even-numbered bulk alkanes. Whether a phase
of the alkane layered crystals is tilted or not, can be understood
on the basis of packing considerations combined with the symmetry
of the molecule. The mirror plane of the odd numbered alkanes
perpendicular to the long axis of the molecule calls for no tilt,
whereas the inversion symmetry of the even alkanes tolerates
finite tilt angles. Disorder of almost any kind makes odd and even
molecules appear equivalent. This is in particular obvious for the
case of orientational disorder with respect to rotations about the
molecular axis. If disorder destroys the lamellar arrangement, as
appears to be the case in pores, the question of tilt is
irrelevant.

In summary, we have shown that pore confinement stabilizes the
rotator mesophases, such that they appear even in C14 and C16
where they are absent in the bulk state.

\begin{acknowledgments}
This work has been supported by the Sonderforschungsbereich 277,
Saarbr\"ucken.
\end{acknowledgments}


\end{document}